\RequirePackage{fix-cm}
\documentclass{svjour3}
\usepackage [latin1]{inputenc}
\usepackage{graphicx,mathptmx,latexsym,amsmath,amssymb,bm}
\smartqed
\journalname{Transport in Porous Media}

\begin{document}
%--------------------------------------------------------------------
\title{Linearity of the co-moving velocity}
%--------------------------------------------------------------------
\titlerunning{Co-moving velocity}
\author{Alex Hansen}

\institute{PoreLab, Department of Physics, Norwegian University of Science and Technology, NO--7491 Trondheim, Norway.\\
\email{alex.hansen@ntnu.no}\\
}

\date{Received: date / Accepted: date}
\maketitle

\begin{abstract}
The co-moving velocity is a new variable in the description of immiscible two-phase flow
in porous media. It is the saturation-weighted average over the derivatives of the seepage 
velocities of the two immiscible fluids with respect to saturation. Based on analysis of
relative permeability data and computational modeling, it has been proposed that the
co-moving velocity is linear when plotted against the derivative of the average seepage 
velocity with respect to the saturation, the flow derivative. I show here that it is 
enough to demand that the co-moving velocity is characterized by an additive parameter 
in addition to the flow derivative to be linear.  This has profound consequences for relative
permeability theory as it leads to a differential equation relating the two relative 
permeabilities describing the flow. I present this equation together with two solutions. 
\end{abstract}
%--------------------------------------------------------------------
\maketitle

%%%%%%%%%%%%%%%%%%%%%%%%%%%%%%%%%%%%%%%%%%%%%%%%%%%%%%%%%
\section{Introduction}
\label{intro}
%%%%%%%%%%%%%%%%%%%%%%%%%%%%%%%%%%%%%%%%%%%%%%%%%%%%%%%%%

The description of immiscible and incompressible two-phase flow in porous media at scales 
large enough so that the porous medium may be seen as a continuum, has been a challenge 
for a long time. In 1936 Wyckoff and Botset \cite{wyckoff1936flow} proposed a generalization
of the Darcy equation \cite{darcy1856fontaines} by splitting it into two independent 
Darcy-type equations, one for each fluid, based on the idea that each fluid will experience, is the matrix and the other fluid together acting as an effective porous matrix.  
This reduces the effective permeability for each fluid by a factor, the relative permeability. It is then assumed that the two relative permeabilities depend only on the saturation.  If I assume steady-state flow and that there are no continuum-scale gradients in the saturation, the theory is complete as described. This theory has a dominating position in all practical calculations. 

The assumption that the relative permeabilities only depends on the saturation is the weak part of the theory.  It is fairly accurate in a window of capillary numbers.  However, outside this window, the pressure gradient enters and the theory loses its predictive power. For example Tallakstad et al.\ found that the average flow rate through a model porous medium over a rather wide range of capillary numbers depends on the pressure drop across it to a power around two \cite{tallakstad2009steady,tallakstad2009steady2}. In their experiment, they controlled the flow rates of the two immiscible fluids entering the 
porous medium, measuring the pressure drop and the saturation.  But how to construct two generalized relative permeabilities from this information?  

Hansen et al.\ \cite{hansen2018relations,feder2022physics} recognized that the 
problem is that knowing the flow velocity of each fluid makes it possible to find the 
combined average flow velocity, but the opposite is not.  The combined average flow velocity 
must be accompanied by a second independent velocity in order to calculate the two 
separate fluid velocities from the average flow velocity. This would be the 
{\it co-moving\/} velocity.  Roy et al.\ \cite{roy2022co} measured the co-moving 
velocity from relative permeability data using multiple sources and calculated it in a dynamic pore network model, finding that it would produce a straight line when plotted against the derivative of the average flow velocity with respect to the saturation --- 
the {\it flow derivative\/} in the terminology of Hansen et al.\ 
\cite{hansen2023statistical} --- necessitating only two parameters, one setting the scale and one determining the slope. They observed this linear behavior in the co-moving velocity plots generated from the relative permeability data,   When the dynamic pore 
network model went outside the limits of relative permeability theory and into the 
non-linear flow regime first observed by Tallakstad et al.\ 
\cite{tallakstad2009steady,tallakstad2009steady2}, the linearity of the co-moving velocity was still in place, but the two parameters would now depended on the pressure gradient.  
Hansen et al.\ \cite{hansen2023statistical} constructed a statistical mechanics
for immiscible two-phase flow in porous media under steady-state conditions that would explain  why the derivative of the average flow velocity with respect to the saturation would be the natural variable for the co-moving velocity.  Pedersen and Hansen
\cite{pedersen2023parameterizations} then presented an interpretation of the co-moving velocity by exploring the geometry of the space of variables describing the flow. 

The last paper so far addressing the co-moving velocity is Alzubaidi et al.\ 
\cite{alzubaidi2023impact} who measured it for a wide range of contact angles 
using lattice Boltzmann computations on the reconstructed pore spaces of 
North Sea Sandstone and Bentheimer sandstone, but in the linear flow regime. In addition, 
they used experimental flow data on Bentheimer sandstone \cite{zou2018experimental}.  
Their surprising conclusion was that the parameters characterizing the co-moving velocity 
are insensitive to the contact angle.     

Three mysteries can therefore be identified concerning the co-moving velocity: 1.\ why is
it linear in the flow derivative and 2.\ why is it still linear when the average
flow rate becomes non-linear, and 3.\ why is the value of the parameters insensitive to the 
contact angle, at least in the linear regime.  

The aim of this note is to address the first and second of these points.  Is the linearity a 
result of some underlying mechanism that has to do with the detailed physics of immiscible 
two-phase flow, or is it rooted in the mathematical structure of the problem \cite{pedersen2023parameterizations}?  The robustness of the linearity (point two above) hints at the latter as the linear and the non-linear flow regimes are quite different.  
It is this viewpoint I will take in this note.  

The linearity of the co-moving velocity when plotted against the flow derivative is a strong
statement that has profound consequences.  It is the aim of this paper to find a necessary
and sufficient criterion for linearity. 
 
I will claim that it is sufficient to demand that two parameters with dimensions of 
velocity are necessary to characterize the seepage velocity of each fluid, $v_w$ and $v_n$,
where the subscripts refer to the wetting and non-wetting fluids. This does not exclude 
additional dimensionless parameters.  Fix the pressure gradient and do a wetting saturation 
sweep.  When the wetting saturation is at a minimum, which I will assume to be for $S_w=0$, 
the wetting seepage velocity defines the initial level $v_w^0$ (which may be zero). Likewise, when the non-wetting saturation is a minimum, which I will assume to be for $S_n=1-S_w=0$, 
the non-wetting fluid will be flowing at its initial level $v_n^0$ (which may be zero). 
The second parameter I will call the range.  It is for the wetting fluid the difference 
between seepage velocity  at $S_w=1$ minus $v_w^0$.  I will call it $\Delta v_w$.  Likewise, 
I define $\Delta v_n$ for the non-wetting seepage velocity.  The two seepage velocities may 
then be written
\begin{eqnarray}
v_w=v_w^0+\Delta v_w u_w(S_w)\;,\label{eq-1-1}\\
v_n=v_n^0+\Delta v_n u_n(S_n)\;,\label{eq-1-2}
\end{eqnarray}
where $u_w(S_w)$ and $u_n(S_n)$ are dimensionless functions of the saturation. Note that the two 
parameters $v_w^0$ and $v_n^0$ are {\it additive.\/} In the flow regime where one may define 
the relative permeabilities $k_{rw}(S_w)$ and $k_{rn}(S_n)$, I have
\begin{eqnarray}
u_w(S_w)=\frac{k_{rw}(S_w)}{S_w}\;,\label{eq-1-3}\\
u_n(S_n)=\frac{k_{rn}(S_n)}{S_n}\;.\label{eq-1-4}
\end{eqnarray}
As I am working with the seepage velocities and not the Darcy velocities, I need to divide the
relative permeabilities by the saturation. 

I will then demonstrate that two parameters with dimensions of velocity being necessary to 
characterize the seepage velocity of each fluid lead to one parameter with dimension of
velocity characterizing the co-moving velocity.  

A necessary and sufficient criterion for the co-moving being linear in the flow derivative
is then that the parameter with dimension of velocity is additive.

In the next section, I review the central concepts surrounding the co-moving velocity. 
I provide a concrete example in order to illustrate the concepts in a concrete way,
namely that of Corey relative permeabilities with indices both equal to two.  I then
go on to analyse the average seepage velocity in Section \ref{average}. By using the Euler
homogeneity theorem combined with the assumption that two parameters with dimensions of 
velocity in addition to non-dimensional parameters describe the average seepage velocity, I 
derive a general functional form for it. I continue with the example from Section \ref{central}
to illustrate the manipulations. In Section \ref{co-moving} I analyse the co-moving 
velocity using the method used for the average seepage velocity.  Having two parameters with dimensions 
of velocity needed for the average seepage velocities, this leads to only one dimensional 
parameter being needed for the co-moving velocity.  Assuming then that this parameter is an 
added constant, a mild assumption, the Euler homogeneity theorem forces the co-moving velocity 
to be linear. I also here use the previously introduced concrete example to illustrate the 
theory. In Section \ref{relperm} I explore the consequences of this linearity for relative permeability theory.  In particular, I present an equation combining the two relative permeabilities and provide two solutions of the equation.  Finally, I conclude in Section \ref{conclusion}.

%%%%%%%%%%%%%%%%%%%%%%%%%%%%%%%%%%%%%%%%%%%%%%%%%%%%%%%%%
\section{Central concepts}
\label{central}
%%%%%%%%%%%%%%%%%%%%%%%%%%%%%%%%%%%%%%%%%%%%%%%%%%%%%%%%%

I assume a porous medium sample having an area $A$ transversal to the flow direction and a
porosity $\phi$.  The sample is uniform. I inject under steady state conditions a volumetric flow rate $Q_w$ of the (more) wetting fluid and a volumetric flow rate $Q_n$ of the (less) wetting fluid. I assume the fluids to be well mixed in the sample so that there are no sample-scale gradients in the saturation.  The total average flow rate is 
\begin{equation}
\label{eq-2-1}
Q_t=Q_w+Q_n\;.
\end{equation}

I now consider a cut through the porous sample orthogonal to the flow direction.  It has an
area $A$. The sub-area of the transversal area $A$ that cuts through the pores is $A_p=\phi A$.
The sub-area of the pore area $A$ that cuts through the wetting fluid is $A_w=S_wA_p$ and the 
sub-area that cuts through the less wetting fluid is $A_n=S_nA_p$.

I can then define the seepage velocities.  The seepage velocity of the wetting fluid is
\begin{equation}
\label{eq-2-2}
v_w=\frac{Q_w}{A_w}\;,
\end{equation}
and the less wetting fluid,
\begin{equation}
\label{eq-2-3}
v_n=\frac{Q_n}{A_n}\;.
\end{equation}
The combined average seepage velocity of the fluids is given by
\begin{equation}
\label{eq-2-4}
v_t=\frac{Q_t}{A_p}=S_wv_w+S_nv_n\;,
\end{equation}
where I in the last expression have use equations (\ref{eq-2-1}) -- (\ref{eq-2-3}).

The co-moving velocity is defined as \cite{hansen2018relations} 
\begin{equation}
\label{eq-2-5}
v_m=S_w\left(\frac{\partial v_w}{\partial S_w}\right)_p+S_n\left(\frac{\partial v_n}{\partial S_w}\right)_p\;,
\end{equation}
where the subscript $p$ means keeping the pressure gradient $p$ fixed. 
By taking the derivative of equation (\ref{eq-2-4}), I find that the co-moving 
velocity may be written
\begin{equation}
\label{eq-2-6}
v_m=\left(\frac{\partial v_t}{\partial S_w}\right)_p+v_n-v_w=v_t'+v_n-v_n\;,
\end{equation}
where I have defined $v_t`=(\partial v_t/\partial S_w)_p$.

Equations (\ref{eq-2-4}) and (\ref{eq-2-5}) constitute the mapping $(v_w,v_n)\to (v_t,v_m)$.
The inverse mapping, $(v_t,v_m)\to (v_w,v_n)$, is 
\begin{eqnarray}
v_w=v_t+S_n(v_t'-v_m)\;,\label{eq-2-7}\\
v_n=v_t-S_w(v_t'-v_m)\;.\label{eq-2-8}
\end{eqnarray}

Roy et al.\ \cite{roy2022co} and Alzubaidi et al.\ \cite{alzubaidi2023impact} observed
the co-moving velocity being linear when plotted against $v_t'$,
\begin{equation}
\label{eq-2-9}
v_m=a(p)+b(p)v_t'\;.
\end{equation}
It is important for the derivation in the next section to note that $v_t'$ is a 
{\it control parameter\/} in this expression in accordance with 
\cite{hansen2023statistical}.  Hence, it does not depend on anything, but other 
functions will depend on it. 

I refrain in the following from writing the pressure gradient explicitly as it does
not enter the discussion explicitly before Section \ref{relperm}.

%%%%%%%%%%%%%%%%%%%%%%%%%%%%%%%%%%%%%%%%%%%%%%%%%%%%%%%%%
\subsection{An example}
\label{example-2-1}
%%%%%%%%%%%%%%%%%%%%%%%%%%%%%%%%%%%%%%%%%%%%%%%%%%%%%%%%%

In order to illustrate these manipulations, I will consider a 
concrete example. Suppose
\begin{eqnarray}
v_w&=&v_w^0+\Delta v_w S_w\;,\label{eq-2-1-1}\\
v_n&=&v_n^0+\Delta v_n S_n\;.\label{eq-2-1-2}
\end{eqnarray}
Hence, I have set $u_w(S_w)=S_w$ and $u_n(S_n)=S_n$ in equations (\ref{eq-1-1}) and
(\ref{eq-1-2}). According to equations (\ref{eq-1-3}) and (\ref{eq-1-4}) this 
corresponds to using the Corey relative permeabilities with indices both equal to two. 
The parameters $v_w^0$, $v_n^0$, $\Delta v_w$, and $\Delta v_n$ all have dimensions of velocity.  

The average seepage velocity, equation
(\ref{eq-2-4}), is then
\begin{equation}
\label{eq-2-1-5}
v_t=v_t^0-4\Delta v_t\ C\ S_w(1-C\ S_w)\;,
\end{equation}
where
\begin{equation}
\label{eq-2-1-6}
v_t^0=v_n^0+\Delta v_n\;,
\end{equation}
and
\begin{equation}
\label{eq-2-1-7}
\Delta v_t=\frac{(v_n^0-v_w^0+2\Delta v_n)^2}{4(\Delta v_w+\Delta v_n)}
\end{equation}
have dimension of velocity. The parameter 
\begin{equation}
\label{eq-2-1-8}
C= \frac{\Delta v_w+\Delta v_n}{v_n^0-v_w^0+2\Delta v_n}
\end{equation}
is non-dimensional.   

The flow derivative may be written
\begin{equation}
\label{eq-2-1-9}
v_t'=v_w^0-v_n^0-2\Delta v_n+2(\Delta v_w+\Delta v_n)S_w=\Delta v_t\ 4C\ (2C\ S_w-1)\;.
\end{equation}
I invert this expression with respect to $S_w$, finding 
\begin{equation}
\label{eq-2-1-10}
S_w=\frac{1}{2C}\left(1+\ \frac{4C\ v_t'}{\Delta v_t}\right)\;.
\end{equation}
I express $v_t$ in terms of $v_t'$ to find
\begin{equation}
\label{eq-2-1-11}
v_t=v_t^0-\Delta v_t+\frac{v_t'^2}{16C^2\Delta v_t}\;.
\end{equation}

Combining equation (\ref{eq-2-5}) with equations (\ref{eq-2-1-1}) and (\ref{eq-2-1-2}) gives
the co-moving velocity
\begin{equation}
\label{eq-2-1-12}
v_m=-\Delta v_n+(\Delta v_w+\Delta v_n)S_w\;.
\end{equation}
which becomes
\begin{equation}
\label{eq-2-1-13}
v_m=\frac{v_n^0-v_w^0}{2}+\frac{v_t'}{2}\;
\end{equation}
when expressed in terms of the variable $v_t'$.  With reference to equation (\ref{eq-2-9}), I 
find that
\begin{eqnarray}
a&=&\frac{v_n^0-v_w^0}{2}\;,\label{eq-2-1-14}\\
b&=&\frac{1}{2}\;.\label{eq-2-1-15}
\end{eqnarray}
Parameter $a$ has units of velocity whereas $b$ is non-dimensional. 

%%%%%%%%%%%%%%%%%%%%%%%%%%%%%%%%%%%%%%%%%%%%%%%%%%%%%%%%%
\section{Dimensional analysis of the average velocity}
\label{average}
%%%%%%%%%%%%%%%%%%%%%%%%%%%%%%%%%%%%%%%%%%%%%%%%%%%%%%%%%

My assumption is that the seepage velocity of the two fluids $v_w$ and $v_n$ each need 
two parameters with dimensions of velocity, level and range, see equations (\ref{eq-1-1})
and (\ref{eq-1-2}). Combining equation (\ref{eq-2-4}) with these two equations will 
generate two parameters with dimensions of velocity $v_t^0$ and $\Delta v_t$ that
will be functions of the four parameters $v_w^0$, $v_n^0$, $\Delta v_w$, and $\Delta v_n$. 

The average seepage velocity $v_t$ must be invariant under change of units.  This means that 
if I divide the two parameters $v_t^0$ and $\Delta v_t$, and the flow derivative $v_t'$ by a 
velocity scale $v_0$, the relation between them stays unchanged. In other words, they obey
\begin{equation}
\label{eq-3-1}
\frac{v_t}{v_0}=v_t\left(\frac{v_t^0}{v_0},\frac{\Delta v_t}{v_0},\frac{v_t'}{v_0}\right)\;.
\end{equation}
This defines $v_t$ as an Euler homogeneous function of order one in $(v_t^0,\Delta v_t, v_t')$.  
I take the derivative with respect to $1/v_0$ and set it equal to one.  This gives 
\begin{equation}
\label{eq-3-2}
v_t=\left(\frac{\partial v_t}{\partial v_t^0}\right)_{\Delta v_t,v_t'} v_t^0+
    \left(\frac{\partial v_t}{\partial \Delta v_t}\right)_{v_t^0,v_t'} \Delta v_t+
    \left(\frac{\partial v_t}{\partial v_t'}\right)_{v_t^0\Delta v_t} v_t'\;.
\end{equation}
I set
\begin{equation}
\label{eq-3-3}
\left(\frac{\partial v_t}{\partial \Delta v_t}\right)_{v_t^0,v_t'}=f\;,
\end{equation}
and
\begin{equation}
\label{eq-3-4}
\left(\frac{\partial v_t}{\partial v_t'}\right)_{v_t^0\Delta v_t}=g\;,
\end{equation}
so that equation (\ref{eq-3-2}) becomes
\begin{equation}
\label{eq-3-5}
v_t=\left(\frac{\partial v_t}{\partial v_t^0}\right)_{\Delta v_t,v_t'} v_t^0+f \Delta v_t+g v_t'\;.
\end{equation}
Both $f$ and $g$ are homogeneous functions of order zero in $(v_t^0,\Delta v_t, v_t')$.

The parameter $v_t^0$ sets the level.  It is therefore an added constant.  This means that
\begin{equation}
\label{eq-3-6}
\left(\frac{\partial v_t}{\partial v_t^0}\right)_{\Delta v_t,v_t'}=1\;.
\end{equation}
From equation (\ref{eq-3-5}) we have that
\begin{eqnarray}
\label{eq-3-7}
\left(\frac{\partial v_t}{\partial v_t^0}\right)_{\Delta v_t,v_t'}&=&\left(\frac{\partial}{\partial v_t^0}\right)_{\Delta v_t,v_t'}
\left[v_t^0+f \Delta v_t+gv_t'\right]\nonumber\\
&=&1+\left(\frac{\partial f}{\partial v_t^0}\right)_{\Delta v_t,v_t'}\Delta v_t +\left(\frac{\partial g}{\partial v_t^0}\right)_{\Delta v_t,v_t'}v_t'\;,
\end{eqnarray}
where I have used equation (\ref{eq-3-6}). This gives 
\begin{equation}
\label{eq-3-8}
\left(\frac{\partial f}{\partial v_t^0}\right)_{\Delta v_t,v_t'}\Delta v_t +\left(\frac{\partial g}{\partial v_t^0}\right)_{\Delta v_t,v_t'}v_t'=0\;,
\end{equation}
for all choices of $\Delta v_t$ and $v_t'$.  I conclude that neither $f$ nor $g$ depends on $v_t^0$, and we may write $v_t$ as
\begin{equation}
\label{eq-3-9}
v_t=v_t^0+f\left(\frac{v_t'}{\Delta v_t}\right)\Delta v_t+g\left(\frac{v_t'}{\Delta v_t}\right)v_t'\;,
\end{equation}
where $f$ and $g$ both are functions of the dimensionless single variable $(v_t'/\Delta v_t)$. 

I note that the dimensional analysis just performed does not generate any constraints on the two functions $f$ and $g$ apart from both being functions of the non-dimensional variable $(v_t'/\Delta v_t)$. 
 
%%%%%%%%%%%%%%%%%%%%%%%%%%%%%%%%%%%%%%%%%%%%%%%%%%%%%%%%%
\subsection{The example: average velocity}
\label{example-3-1}
%%%%%%%%%%%%%%%%%%%%%%%%%%%%%%%%%%%%%%%%%%%%%%%%%%%%%%%%%

Here I calculate $f$and $g$, equations (\ref{eq-3-3}) and (\ref{eq-3-4}) for the example given 
in Section \ref{example-2-1}. By using equation (\ref{eq-2-1-11}) I find
\begin{equation}
\label{eq-3-1-1}
f\left(\frac{v_t'}{\Delta v_t}\right)=-1-\frac{1}{16C^2}\ \left(\frac{v_t'}{\Delta v_t}\right)^2\;,
\end{equation}
and
\begin{equation}
\label{eq-3-1-2}
g\left(\frac{v_t'}{\Delta v_t}\right)=\frac{1}{8C^2}\ \left(\frac{v_t'}{\Delta v_t}\right)\;.
\end{equation}
Combining these two expressions with equation (\ref{eq-3-5}) recovers equation (\ref{eq-2-1-11}).
I note that even if I set $v_w^0=v_n^0=0$, $v_t^0=\Delta v_n$ is still non-zero.

%%%%%%%%%%%%%%%%%%%%%%%%%%%%%%%%%%%%%%%%%%%%%%%%%%%%%%%%%
\section{Dimensional analysis of the co-moving velocity}
\label{co-moving}
%%%%%%%%%%%%%%%%%%%%%%%%%%%%%%%%%%%%%%%%%%%%%%%%%%%%%%%%%

The three velocities, $v_w$, $v_n$ and $v_t$ each contain two dimensional parameters of which one is an 
added constant.  The definition of the co-moving velocity (\ref{eq-2-5}) contains 
the derivatives of $v_w$ and $v_n$ with respect to the saturation. Hence, the two parameters 
$v_w^0$ and $v_n^0$, defined in equations (\ref{eq-1-1}) and (\ref{eq-1-2}) disappear 
when the co-moving velocity is expressed as a function of the saturation $S_w$.  Only
the parameters $\Delta v_w$ and $\Delta_n$ appear, i.e.,
\begin{equation}
\label{eq-4-1}
v_m=v_m(\Delta v_w,\Delta v_n, S_w)\;.
\end{equation}
I now express the co-moving velocity using the flow derivative $v_t'$ rather than
the saturation $S_w$. As $v_t'$ has the dimensions of velocity, there will now only be one
parameter with this dimensionality, the parameter $a$.  Hence, we have 
\begin{equation}
\label{eq-4-2}
v_m=v_m(a,v_t')\;.
\end{equation}

The co-moving velocity must be invariant with respect to a change of 
velocity scale $v_0$.  This was e.g., used in the analysis in Roy et al.\ \cite{roy2022co}, 
as it was the basis for using the relative permeability data. This means that $v_m$ 
obeys the scaling relation 
\begin{equation}
\label{eq-4-3}
\frac{v_m(a,v_t')}{v_0}=v_m\left(\frac{a}{v_0},\frac{v_t'}{v_0}\right)\;.
\end{equation}
Hence, $v_m$ is also an Euler homogeneous function of order one in the parameter $a$
and the variable $v_t'$.  Taking the derivative with respect to $1/v_0$ and setting $v_0=1$ gives
\begin{equation}
\label{eq-4-4}
v_m(a,v_t')=\left(\frac{\partial v_m}{\partial a}\right)_{v_t'} a + \left(\frac{\partial v_m}{\partial v_t'}\right)_{a} v_t'\;.
\end{equation}

I define
\begin{eqnarray}
\left(\frac{\partial v_m}{\partial a}\right)_{v_t'}=c\left(\frac{v_t'}{a}\right)\;,\label{eq-4-5}\\
\left(\frac{\partial v_m}{\partial v_t'}\right)_{a}=b\left(\frac{v_t'}{a}\right)\;,\label{eq-4-6}
\end{eqnarray}
where $c(v_t/a)$ and $b(v_t'/a)$ are homogeneous functions of degree zero.  Hence, 
\begin{equation}
\label{eq-4-7}
v_m(a,v_t')=c\left(\frac{v_t'}{a}\right) a + b\left(\frac{v_t'}{a}\right) v_t'\;.
\end{equation}

I must have that
\begin{equation}
\label{eq-4-8}
\left(\frac{\partial c}{\partial v_t}\right)_{a}
=\frac{\partial^2 v_m}{\partial v_t \partial a}=\frac{\partial^2 v_m}{\partial a \partial v_t'}
=\left(\frac{\partial b}{\partial a}\right)_{v_t'}\;.
\end{equation}
I now introduce a variable $x$ defined as
\begin{equation}
\label{eq-4-9}
x=\frac{v_t'}{a}\;,
\end{equation}
so that equation (\ref{eq-4-8}) is reduced to the equation
\begin{equation}
\label{eq-4-10}
\frac{dc(x)}{dx}+x\ \frac{db(x)}{dx}=0\;.
\end{equation}

I note that if the parameter $a$ is additive, the function $c$ is  
\begin{equation}
\label{eq-4-11}
c(x)=1\;,
\end{equation}
i.e., a constant. From equation (\ref{eq-4-10}) I then have that the function $b(x)$ is also a constant. 
Combining this result with equation (\ref{eq-3-7}) gives
the final result,
\begin{equation}
\label{eq-4-12}
v_m(a,v_t')=a+b\ v_t'\;.
\end{equation}
Hence, {\it the co-moving velocity $v_m$ is linear in $v_t'$ if and only if $a$ is an additive
constant.\/}

The assumption that $a$ is an additive constant makes it possible to identify it.  From equation (\ref{eq-2-6}) 
I have that  
\begin{equation}
\label{eq-4-13}
a=v_m(a,0)=[v_n-v_w]_{v_t'=0}\;.
\end{equation}

This identification is only possible if $v_t'=0$ falls within the physical range of the
parameters; in other words, $v_t$ has a minimum in the range $0\le S_w\le 1$.  If
there is no such minimum, it is still possible to define $a$, but it will not 
have the simple physical interpretation given in equation (\ref{eq-4-13}).

%%%%%%%%%%%%%%%%%%%%%%%%%%%%%%%%%%%%%%%%%%%%%%%%%%%%%%%%%
\subsection{The example: co-moving velocity}
\label{example-4-1}
%%%%%%%%%%%%%%%%%%%%%%%%%%%%%%%%%%%%%%%%%%%%%%%%%%%%%%%%%

The co-moving velocity in the example is given in equation
(\ref{eq-2-1-12}). In order to calculate $a$ according to equation (\ref{eq-4-9}), 
I set $v_t'=0$ in the equation (\ref{eq-2-1-10})
giving the corresponding saturation $S_w=1/2C$. I use the equations 
for $v_w$ and $v_n$, (\ref{eq-2-1-1}) and (\ref{eq-2-1-2}) together
with the definition of $C$, equation (\ref{eq-2-1-8}),
\begin{equation}
\label{eq-4-1-1}
a=[v_n-v_w]_{v_t'=0}=v_n^0-v_w^0+\Delta v_n -\frac{1}{2C} (\Delta v_w+\Delta v_n)
=\frac{v_n^0-v_w^0}{2}\;.
\end{equation} 
We see that $v_w^0$ and $v_n^0$ appear in the expression in spite of equation (\ref{eq-4-1}).  
This is so since switching from $S_w$ to $v_t'$ as control variable reintroduces them.

%%%%%%%%%%%%%%%%%%%%%%%%%%%%%%%%%%%%%%%%%%%%%%%%%%%%%%%%%
\section{Consequences for relative permeability theory}
\label{relperm}
%%%%%%%%%%%%%%%%%%%%%%%%%%%%%%%%%%%%%%%%%%%%%%%%%%%%%%%%%

The generalized Darcy equations that forms the core of relative permeability theory are
\begin{eqnarray}
v_w=-\frac{Kk_{rw}}{\mu_w\phi S_w}\ p\;,\label{eq-5-1}\\
v_n=-\frac{Kk_{rn}}{\mu_n\phi S_n}\ p\;,\label{eq-5-2}
\end{eqnarray}
where $\mu_w$ and $\mu_n$ are viscosities, $K$ the permeability and $k_{rw}(S_w)$ and
$k_{rn}(S_w)$ are the relative permeabilities.  The saturations appear in the denominators
since $v_w$ and $v_n$ are seepage velocities.  

Following Roy et al.\ \cite{roy2022co}, I define a velocity scale
\begin{equation}
\label{eq-5-3}
v_0=-\frac{K}{\mu_w\phi}\ p\;,
\end{equation}
which is the seepage velocity of the wetting fluid at saturation $S_w=1$.  I may then write
the generalized Darcy equations as
\begin{eqnarray}
v_w&=&\frac{k_{rw}}{S_w}\ v_0\;,\label{eq-5-4}\\
v_n&=&\frac{k_{rn}}{S_n}\ M v_0\;,\label{eq-5-5}
\end{eqnarray}
where I have defined the viscosity ratio $M=\mu_w/\mu_n$. The average velocity, equation  
(\ref{eq-2-4}), is then given by
\begin{equation}
\label{eq-6-1}
v_t(p,S_w)=\left[k_{rw}(S_w)+M k_{rn}(S_w)\right]v_0(p)\;.
\end{equation}
I now combine equations (\ref{eq-5-4}) and (\ref{eq-5-5}), with the co-moving velocity, 
equation (\ref{eq-2-6}) using equation (\ref{eq-2-9}).  I also make a change in notation by
setting $a\to a v_0$, as in Roy et al.\ \cite{roy2022co}.  The resulting equation is
\begin{equation}
\label{eq-5-6}
(1-b)\frac{d}{dS_w}\left[k_{rw} +Mk_{rn}\right]-\frac{k_{rw}}{S_w}+\frac{Mk_{rn}}{S_n}=a\;.
\end{equation}
A simplified version of this equation was used in Alzubaidi et al.\ 
\cite{alzubaidi2023impact} to determine the non-wetting relative permeability when the  
$k_{rw}$, $a=0$ and $b$ are known.  

It is also possible to find analytic solutions to equation (\ref{eq-5-6}).  I give an example
in the following. Picchi and Battiato \cite{picchi2019relative} suggested the semi-empirical form
(note that their viscosity ratio is the inverse of the definition I use)
\begin{eqnarray}
k_{rw}&=&k_{rw}^0S_w^2\;,\label{eq-5-7}\\
k_{rn}&=&k_{rn}^0S_n^2\left[1+\frac{2S_w}{MS_n}\right]\;,\label{eq-5-8}
\end{eqnarray}
where $k_{rw}^0$ and $k_{rn}^0$ are the end point relative permeabilities. Setting 
\begin{equation}
\label{eq-5-9}
b=\frac{1}{2}\;,
\end{equation} 
and
\begin{equation}
\label{eq-5-10}
a=k_{rn}^0
\end{equation}
solves equation (\ref{eq-5-6}).

By solving for the saturation for which $v_t'=0$, using equations (\ref{eq-5-7}) and 
(\ref{eq-5-8}), I find that 
\begin{equation}
\label{eq-5-11}
[v_n-v_w]_{v_t'=0}=k_{rn}^0\;.
\end{equation}
According to equation (\ref{eq-3-2}) this is equal to $a$, as found solving equation 
(\ref{eq-5-6}) directly, leading to (\ref{eq-5-10}).

The Corey empirical relative permeabilities
\begin{eqnarray}
k_{rw}=k_{rw}^0 S_w^{\mu_w}\;,\label{eq-5-12}\\
k_{rn}=k_{rn}^0 S_n^{\mu_n}\;,\label{eq-5-13}
\end{eqnarray}
are, on the other hand, not solutions to equation 
(\ref{eq-5-6}), except when $\mu_w=\mu_n$, they are solutions, but with 
$a=0$. In that case I find  
\begin{equation}
\label{eq-5-14}
b=1-\frac{1}{\mu_w}\;.
\end{equation}
The example I gave in Section \ref{example-2-1} was based on the Corey relative 
permeabilities with exponents $\mu_w=\mu_n=2$.  The non-zero $a$ that was found, 
equation (\ref{eq-2-1-14}), came from the assumption that $v_w^0$ and $v_n^0$ were non-zero. 
The generalized Darcy equations (\ref{eq-5-1}) and (\ref{eq-5-2}) assumes $v_w^0$ and $v_n^0$
are both zero.

%%%%%%%%%%%%%%%%%%%%%%%%%%%%%%%%%%%%%%%%%%%%%%%%%%%%%%%%%
\section{Discussion and Conclusion}
\label{conclusion}
%%%%%%%%%%%%%%%%%%%%%%%%%%%%%%%%%%%%%%%%%%%%%%%%%%%%%%%%%

I have in this paper shown that the co-moving velocity is of the form given by
equation (\ref{eq-2-9}) given that two conditions are in place: 1.\ the
average seepage velocity is characterized by two parameters with dimensions of 
velocity, allowing for additional dimensionless parameters, and 2.\ the necessity
of an additive velocity scale $a$ in the co-moving velocity. 
Both assumptions are weak and does not hinge on any further restrictions 
on the constitutive law for $v_t$, linear or non-linear.  
Equation (\ref{eq-5-6}), which is the combination of the relative permeability 
constitutive laws (\ref{eq-5-1}) and (\ref{eq-5-2}) with the linear equation for $v_m$, 
(\ref{eq-2-6}), leads to a relation between the two relative permeabilities: knowing one, 
gives the other one.  

The parameter $a=[v_n-v_w]_{v_t'=0}$, equation (\ref{eq-3-2}) may be given a straight-forward
physical interpretation when $v_t=0$ falls within the physical range of the parameters.  
However, the second parameter, $b$ is not.  It typically takes on a value around 0.6--0.8 and 
seems quite insensitive to the contact angle between the fluids and the matrix as shown by 
Alzubaidi et al.\ \cite{alzubaidi2023impact}. It is an open question why.   

%%%%%%%%%%%%%%%%%%%%%%%%%%%%%%%%%%%%%%%%%%%%%%%%%%%%%%%%%

{\sl Declaration} --- The author declares no conflict of interest. 

%%%%%%%%%%%%%%%%%%%%%%%%%%%%%%%%%%%%%%%%%%%%%%%%%%%%%%%%%

{\sl Acknowledgement} --- The author thanks R. T. Armstrong, C.\ F.\ Berg, 
S.\ Berg, E.\ G.\ Flekk{\o}y, J.\ E.\ McClure, B.\ Hafskjold, H.\ Pedersen, 
T.\ Ramstad, S.\ Roy, and S.\ Sinha for countless discussions.  This work was partly supported by the 
Research Council of Norway through its Centres of Excellence funding scheme, project number 262644. 

%%%%%%%%%%%%%%%%%%%%%%%%%%%%%%%%%%%%%%%%%%%%%%%%%%%%%%%%%
\bibliography{references}{}
%%%%%%%%%%%%%%%%%%%%%%%%%%%%%%%%%%%%%%%%%%%%%%%%%%%%%%%%%

\end{document}